# Optical detection of the quantum Hall effect in silicon nanostructures

N.T. Bagraev [1] , L.E. Klyachkin [1]✉ , A.M. Malyarenko [1] , N.I. Rul [2]

[1] Ioffe Institute, St. Petersburg, Russia

[2] Peter the Great St. Petersburg Polytechnic University, St. Petersburg, Russia

✉ leonid.klyachkin@gmail.com

**ABSTRACT**
Electroluminescence spectra of a silicon nanostructure with edge channels covered by chains of dipole centers with negative correlation energy are demonstrated. The presence of such chains provides conditions for nondissipative transport of single charge carriers at high temperatures up to room temperature. Due to the suppression of the electron-electron interactions, the macroscopic quantum phenomena such as Shubnikov–de Haas oscillations and the quantum staircase of Hall resistance are consistent with the positions of the spectral peaks of the detected electroluminescence. The obtained results are considered in the framework of Faraday electromagnetic induction, which indicates that Landau quantization leads to the emergence of induced irradiation similar to Josephson and Andreev generation. Moreover, the detected maxima in the spectral characteristics correspond to odd fractional values of the resistance quantum staircases, while the dips in the electroluminescence spectra are observed at even fractional values of the resistance quantum ladder, which is due to the increased formation of composite bosons and fermions, respectively.

**Funding.** *The work was financed within the framework of the state assignment of the Federal State Unitary Enterprise Ioffe Institute No. FFUG-2024-0039 of the Ministry of Science and Higher Education of the Russian Federation.*

**Citation:** Bagraev NT, Klyachkin LE, Malyarenko AM, Rul' NI. Optical detection of the quantum Hall effect in silicon nanostructures. *Materials Physics and Mechanics*. 2026;54(1): 1–7.
http://dx.doi.org/10.18149/MPM.5412026_1

## Introduction

Currently, macroscopic quantum phenomena in semiconductor nanostructures are predominantly studied at ultra-low temperatures due to the decisive role of electron-electron interactions in suppressing nondissipative transport of single charge carriers [1,2]. Graphene is an exception, yet even in this case, the realization of nondissipative transport conditions at high temperatures requires the application of a strong magnetic field [3].

An emerging breakthrough in solving this problem is associated with advances in technology and fundamental research on topological structures [4–15]. In particular, it has been demonstrated that in ultra-narrow quantum wells with edge channels covered by chains of centers with negative correlation energy (negative U), it is possible to create conditions for nondissipative carrier transport and, consequently, observe the macroscopic quantum phenomena at high temperatures, up to room temperature [16]. The suppression of electron–electron interactions in this case is primarily related to the formation of impurity dipoles due to the dissociation of neutral negative-U centers: $2D^0 \Rightarrow D^+ + D^- + U$, where U is the effective energy that accounts for the compensation of Coulomb repulsion owing to coupling with electron–vibration interactions (EVI) [17–19]. As a result of the negative correlation energy, it is feasible in many systems of impurity and structural centers to form the edge channel shells consisting of chains of negative-U dipole centers, which facilitates the segmentation of edge channels into sections (pixels) containing single charge carriers. Thus, two problems are resolved simultaneously: on one hand, the electron–electron interaction is suppressed, and on the other hand, conditions for nondissipative transport are ensured through energy exchange between a single charge carrier and the negative-U dipole centers of the edge channel shell [20].

In the presence of chains of negative-U dipole centers, nearly nondissipative transport of single charge carriers within the pixels constituting the edge channel is achieved: $h + D^- \Rightarrow D^0$, $D^+ + e \Rightarrow D^0 + h$, $D^- + D^+ + e + h \Rightarrow 2D^0 + h$. Subsequently, hole tunneling along the pixel shell is accompanied by ultrafast formation of negative-U dipole centers: $2D^0 + h \Rightarrow D^- + D^+$. Thus, the presence of negative-U centers promotes the emergence of an energy reservoir within the edge channel shells, enabling weakly nondissipative carrier transport.

In this work, boron centers embedded via diffusion into wafers of monocrystalline silicon (100) under fabrication conditions of the structure in the Hall geometry which is oriented along the [011] axis (Fig. 1) are used as negative-U dipole centers.

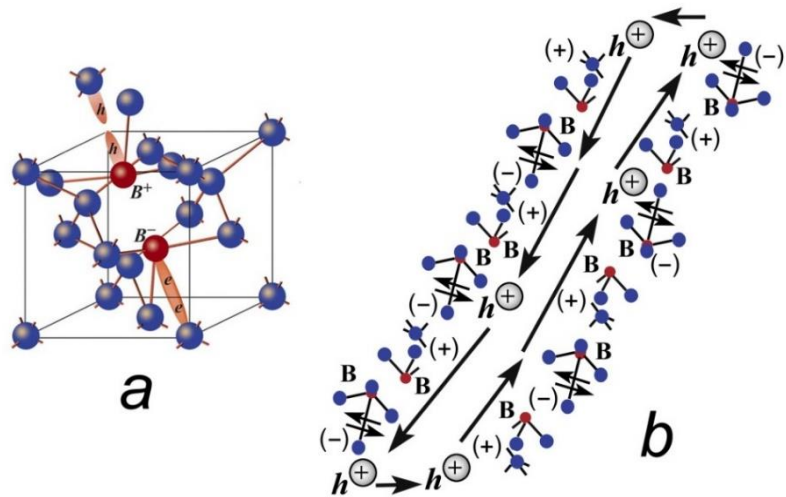

**Fig. 1.** Dipole trigonal boron center ($B^+$–$B^-$) with negative correlation energy (a) and the chains of the dipole boron centers in the δ-barriers confining the ultra‑narrow silicon quantum well and its edge channels (b)

Negative-U shells with similar properties can be obtained not only by embedding a high concentration of impurity centers, but also by fabricating quasi-one-dimensional chains of point and extended defects [21]. For example, low-dimensional structures with similar properties within negative-U shells were realized at the interface of hybrid SiC/Si structures grown by the method of coordinated substitution of atoms [21]. In this case, the role of negative-U dipole center chains is played by crystallographically aligned chains of silicon vacancies as well as paired $C_i - V_{Si}$ centers [21].

It should be noted that the initial reduction in entropy of the semiconductor nanostructure due to the presence of negative-U shells not only allows the observation of macroscopic quantum phenomena at high temperatures but also enables the study of various processes arising from quantum interference of single charge carriers in edge channels. In this case, the pixels act as quantum boxes, and by investigating processes inside them, one can evaluate the relative contributions of equilibrium and non-equilibrium effects to quantum interference. In this work, it is demonstrated through a comparative analysis of the results obtained from studying the electrical and optical versions of the quantum Hall effect (QHE).

**Materials and Method**

The silicon nanostructure was fabricated using planar technology on a monocrystalline silicon (100) surface through preliminary oxidation, subsequent photolithography and gas-phase boron diffusion [22]. Careful optimization of thermal oxidation conditions to achieve an ultra-shallow diffusion profile — under a balance of vacancy and kick-out diffusion mechanisms — enabled the passivation of the boundaries of an ultra-narrow quantum well (2 nm) with boron centers. As a result, an ultra-narrow quantum well was obtained, confined by barriers composed of quasi-one-dimensional boron chains (Fig. 2), spaced 2 nm apart. Despite the high boron concentration ($5 \times 10^{21}$ cm$^{-3}$ according to SIMS data [22]), the edge channels of the quantum well showed no activity in EPR measurements. However, studies of field-dependent magnetization using the Faraday method revealed a significant diamagnetic response, which is exhibited in weak magnetic fields [22]. Analysis of temperature and field dependencies of magnetization led to the proposal of a model in which boron impurity centers passivate the edge channel and are responsible for single-carrier transport. As noted above, the ground state of the impurity shells consists of chains of negative-U dipole boron centers, primarily indicated by the diamagnetic response observed when the nanostructure is placed in an external magnetic field [22].

Since the edge channel consists of pixels containing single charge carriers, conditions are established for the realization of nondissipative transport at high temperatures. In studies of this structure, the following phenomena have been observed:

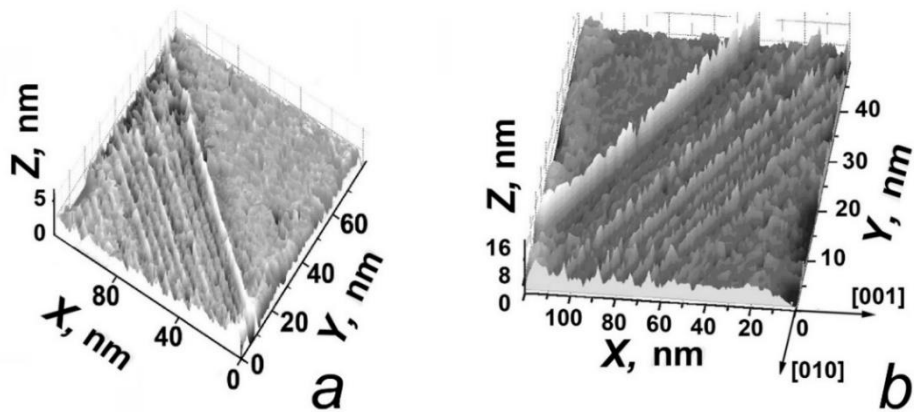

**Fig. 2.** STM image of the silicon nanosandwich (100) surface containing chains of boron dipole centers oriented along the (011) axis

Shubnikov–de Haas and de Haas – van Alphen quantum oscillations, a quantum staircase of Hall resistance, a quantum staircase of conductance [16], as well as electrical [22] and optical [20] versions of the multiple Andreev reflections. In this context, the pixels can be considered as the Andreev molecules [20].

**Results and Discussion**

Figure 3 presents the magnetic field dependencies of the longitudinal ($R_{xx}$) and lateral ($R_{xy}$) resistances of the silicon nanostructure, measured at a temperature of 77 K under a stabilized drain-source current $I_{ds}$ = 10 nA. Since the presence of negative-U boron dipole center chains, which confine the edge channels of the quantum well and divide them into pixels containing individual charge carriers, creates the conditions for dissipationless transport, the characteristics of the Shubnikov–de Haas oscillations and the quantum staircase of Hall resistance are determined by the Landau quantization processes within each individual pixel and directly depend on its geometric dimensions [16,23]. In this case, the emergence of steps in the magnetic field dependence of the Hall resistance can be analyzed within the framework of Faraday's electromagnetic induction [24]:

$$\frac{dE}{d\Phi} = I_{gen}, \quad (1)$$

where $dE$ is the change in carrier energy within the pixel upon a change in magnetic flux ($d\Phi = \Delta B \cdot S$), $S$ is the pixel area and $I_{gen}$ is the induced generation current.

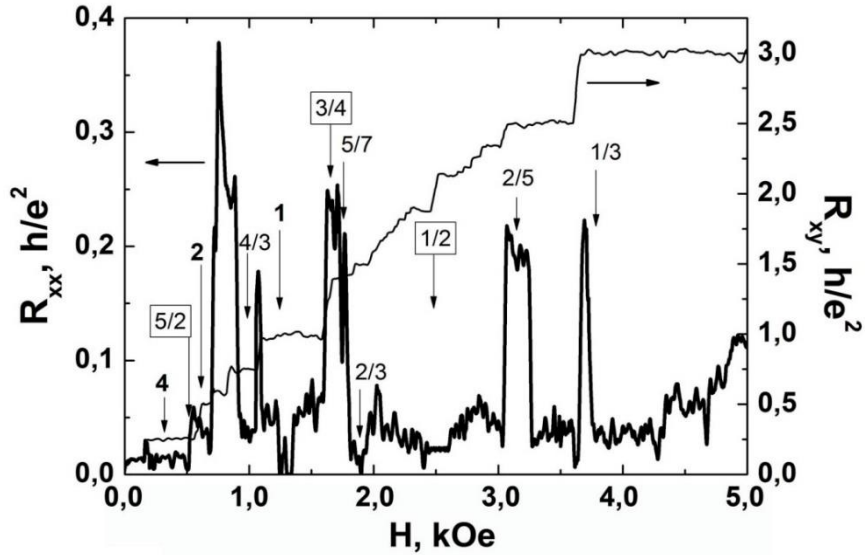

**Fig. 3.** Hall resistance $R_{xy}=V_{xy}/I_{ds}$ and magnetoresistance $R_{xx} = V_{xx}/I_{ds}$ of the silicon nanostructure ($p_{2D}$=3 × 10$^{13}$m$^{-2}$) dependencies on the external magnetic field strength. $T$ = 77 K; $I_{ds}$ = 10 nA

It is appropriate to express the change in magnetic flux as a change in the number of magnetic flux quanta $d\Phi = m\Phi_0$, where $\Phi_0 = h/e$. In turn, the magnitude of the energy change depends on the number of carriers ($n$), $dE = ne \cdot dU$ (for a pixel containing an single charge carrier, $n$ = 1). Thus, the Faraday's relation automatically leads to the step height in the Hall dependence:

$$G = \frac{I_{gen}}{U} = \frac{n}{m} \cdot \frac{e^2}{h}, \quad (2)$$

which makes it possible to describe not only the integral QHE [23] but also the fractional QHE [16].

Approximation of the field-dependent Hall curve allows determination of the carrier density (pixel density), which, according to the dependence in Fig. 3, is $3 \times 10^{13}$ m$^{-2}$. Hence, the pixel dimensions are 2 nm × 16.6 µm, and the number of pixels between the XX contacts is 124.

Since the Josephson and Andreev junctions between the negative-U dipole chains on opposite edges of the pixel stimulate the induced emission, the energy-dependent form of the Faraday's relation can be expressed as:

$$h\nu = eI_{gen}R_N, \qquad (3)$$

where $R_N$ is the load resistance, which in this case corresponds to the quantum resistance of the pixel ($R_N = h/e^2$). Combined with the classical form of the Faraday's relation given above:

$$h\nu = e\left(\frac{h}{e^2}\right)I_{gen} = \left(\frac{h}{e}\right)I_{gen}, h/\Delta E = \tau,$$
$$e/I_{gen} = \tau, I_{gen} = e\Delta E/h, \qquad (4)$$
$$h\nu = e\frac{\Delta E}{h}\Delta BS, \nu = \frac{e\Delta E\Delta BS}{h^2}.$$

Thus, the corresponding wavelengths seem to be selected and construct a scale that aligns with the magnetic field scale in Fig. 3, thereby determining the wavelengths at which features should appear in the electroluminescence spectra corresponding to the positions of the steps in the Hall resistance staircase (Fig. 4).

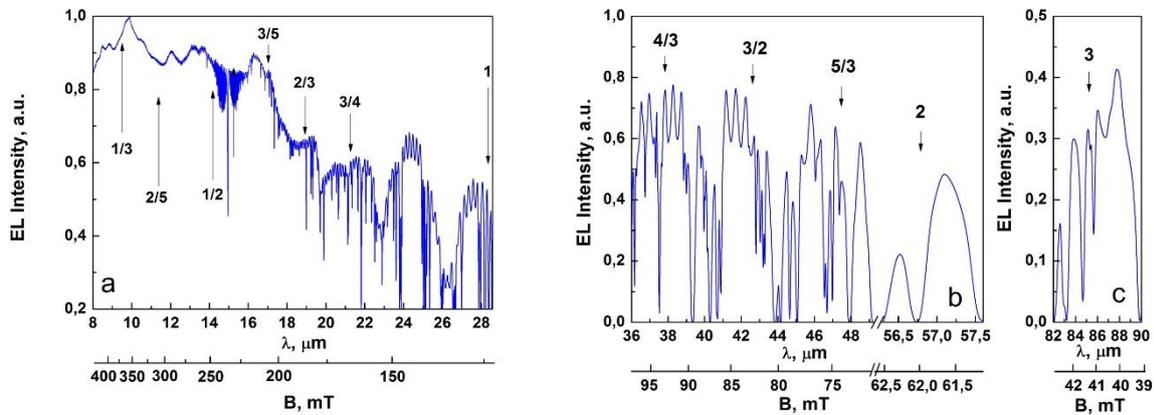

**Fig. 4.** Electroluminescence spectra of the silicon nanostructure in the mid- (a) and far-infrared (b,c) wavelength ranges, showing features corresponding to fractional and integer values of the quantum resistance staircase at *T* = 300 K

It should be noted that the frequency range of the observed electroluminescence extends into the terahertz spectral region, in accordance with the emission mechanism arising from electromagnetic induction. In this case, the proposed version of optical detection of the QHE differs from the first optical detection of the QHE achieved by monitoring changes in interband photoluminescence characteristics in an external magnetic field [25].

Figure 4 presents electroluminescence spectra obtained at *T* = 300 K under a stabilized drain–source current ($I_{ds}$), which induces a magnetic field within the edge channel pixels: $H = \Delta I_{ds}/2r_0$, where $r_0(S/\pi)^{1/2}$ is the effective pixel radius, with $r_0 = 10^{-7}$ m.

The electroluminescence spectra were recorded using the Bruker Vertex 70 infrared Fourier spectrometer. A comparison of the spectral characteristics with the quantum staircase of Hall resistance magnetic field dependencies shows good agreement between the positions of features in the electroluminescence spectra (Fig. 4) and the quantum steps (Fig. 3). It should be noted that the optical peaks corresponding to odd fractional values of the quantum resistance staircase indicate enhanced induced emission — i.e., stimulated generation of composite bosons upon the sequential capture of single magnetic flux quanta. Conversely, the observed dips in the electroluminescence spectra at wavelengths corresponding to even fractional values of the quantum resistance staircase demonstrate enhanced formation of composite fermions, which apparently leads to additional strengthening of the electron–electron interaction and the associated quenching of electroluminescence [26–30].

The consistency between the electrical and optical manifestations of the fractional QHE indicates that Landau quantization gives rise to the induced emission, similar to the Josephson and Andreev generation mechanisms, regardless of how many magnetic flux quanta are involved in the process.

## Conclusion

This work presents the first experimental results demonstrating the possibility of detecting and identifying both the integral and fractional QHEs in the silicon nanostructure with the edge channels confined by the chains of dipole centers exhibiting negative correlation energy, by analysis of features in electroluminescence spectra obtained via infrared Fourier spectroscopy. It is shown that the optical detection of the QHE can be described within the framework of the Faraday electromagnetic induction. The obtained results demonstrate strong agreement between the characteristic behavior of the lateral and longitudinal resistance of the studied silicon nanostructure and the spectra of terahertz electroluminescence arising under the Landau quantization conditions.

## CRediT authorship contribution statement

**Nikolai T. Bagraev** : writing – review & editing, conceptualization; **Leonid E. Klyachkin**:  writing – original draft, investigation; **Anna M. Malyarenko** : investigation, supervision; **Nikolai I. Rul** : investigation, data curation.

## Conflict of interest

The authors declare that they have no conflict of interest.